\documentclass[12pt]{iopart}
\usepackage{iopams}  
\usepackage{graphicx}
\begin{document}
\title{Signature of strong atom-cavity interaction on critical coupling}
\author{Shourya Dutta Gupta$^1$, Subimal Deb$^2$ and S. Dutta Gupta$^{2*}$}
\address{$^1$ Department of Materials and Metallurgical Engineering, IIT Kanpur, Kanpur 208016, India\\
$^2$School of Physics, University of Hyderabad, Hyderabad 500046, India\\
$^*$Corresponding author: sdghyderabad@gmail.com}
\begin{abstract}
We study a critically coupled cavity doped with resonant atoms with metamaterial slabs as mirrors. We show how resonant atom-cavity interaction can lead to a splitting of the critical coupling dip. The results are explained in terms of the frequency and lifetime splitting of the coupled system.
\end{abstract}
\maketitle 
\section{Introduction}
Over the past decade very many exotic applications of negative index materials (NIMs), otherwise known as metamaterials or left-handed materials have been suggested and tested. These range from superlensing to lasing spasers, optical nano-circuits to invisibility cloaks, cavity quantum electrodynamics (QED) to electromagnetically induced transparency \cite{shalaev2007, pendry2000, zheludev2008, engheta2007, xu2009, papasimakis2008}. In most of these applications the doubly negative response of the metamaterials (namely, negative permittivity $\epsilon$ and permeability $\mu$) over a certain range of frequencies has been exploited. In fact, frequency dispersion and absorption have been identified as some of the crucial characteristics of the metamaterials, determining their figure of merit and suitability for device applications \cite{veselago, pendry2004, kinsler2008, nistad2008, sdgjphysb2009}. Only recently it has been recognized that metamaterials, capable of exhibiting negative index, can have interesting applications in a different frequency domain (where it is not truly a negative index material). It was shown that a slab of a metamaterial can exhibit a stop gap (as in one dimensional periodic structures) in the frequency range between the electric and magnetic plasma frequencies \cite{bloemer2005}. Interesting features of such a material in the said frequency domain follows from close to zero real part of the refractive index. Depending on the separation between the plasma frequencies, one thus has a material with broadband absorption features. These properties were exploited to suggest a cavity with metamaterial layers as mirrors, which can be critically coupled to incident radiation, resonant with one or more of the modes of this cavity. Recall that a critically coupled system can absorb almost all the incident light, resulting in near-null reflection as well as transmission. Critical coupling (CC) in layered structures with absorbing polymer films or metal nano composite layers have been studied in detail \cite{tischler2006, sdgol2007, sdgjopa}. It is understood that there are certain advantages for critical coupling in a metamaterial cavity (like flexible tunability, etc.) as compared to the earlier structures. These advantages follow from the fact that now one uses the Fabry-Perot (FP) modes (tunable by changing the cavity free spectral range) without any constraints imposed by the narrowband feature of the polymer or the nano composite absorber. Moreover, there is a field enhancement in the cavity, whenever it is critically coupled. In fact, these are the special features one looks for to study the strong interaction regime of cavity QED. Recall that in the weak interaction regime, the photon emitted by the excited atom can escape the cavity, before it can interact back with the atom. In case of strong interaction, due to high finesse of the cavity, the photon has a larger life time, and there is a periodic exchange of energy between the atom and the cavity mode. It is thus interesting to study atom-cavity interaction in the context of a critically coupled cavity. In this paper, we study the manifestations of this interaction when the intracavity medium is doped with atoms resonant with the critical coupling frequency. We show that like in usual vacuum field Rabi splitting \cite{carmichaelnato}, here also we have the splitting of the critical coupling dip into two, one corresponding to the symmetric mode and the other to the anti-symmetric mode.  We solve the dispersion relation for obtaining the corresponding branches for both the real and the imaginary parts of the frequency. The real part gives the approximate location of the modes, while the imaginary part gives the corresponding decay rates. Finally, we study the field distributions corresponding to the excitation of these modes.

The structure of the paper is as follows. In Section 2, we define our system and formulate the problem. Section 3 is devoted to numerical results pertaining to the frequency splitting in the total scattering data and also the roots of the dispersion equation. Finally, in Conclusions, we summarize the main results of this paper.

\section{Formulation}
\begin{figure}\centering
\includegraphics[width=6cm]{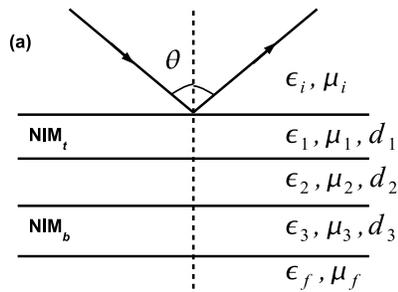}
\caption{Schematic of the layered structure.}\label{fig:scheme}
\end{figure}
Consider the system shown in figure \ref{fig:scheme} consisting of a spacer layer between two slabs of metamaterials. Let the permittivity and permeability of the negative index materials be given by a Lorentz type response as follows \cite{shelby2001}:
\begin{equation}
\begin{array}{lll}
\epsilon(f)&=& 1-\frac{f_{ep}^2-f_{eo}^2}{f^2-f_{eo}^2+i\gamma f}~,\\
\mu(f)     &=& 1-\frac{f_{mp}^2-f_{mo}^2}{f^2-f_{mo}^2+i\gamma f}~,
\end{array}\label{eq:epsmu}
\end{equation}
where $f_{ep}$ ($f_{mp}$) is the electric (magnetic) plasma frequency, $f_{eo}$ ($f_{mo}$) is the electric (magnetic) resonance frequency and $\gamma$ is the decay rate (we assumed the same decay rate for both electric and magnetic resonances). A slab of such a metamaterial was shown to exhibit a stopgap (like in one dimensional photonic band gap structures) between its electric and magnetic plasma frequencies, characterized by  mostly imaginary refractive index \cite{bloemer2005}. We will refer to such metamaterials still as NIMs, though we consider a frequency domain where it does not possess negative refraction. We use two slabs of such NIMs with distinct values of the plasma frequencies such that the stop band of one (NIM$_t$, see figure \ref{fig:scheme}) lies well  inside that of the other (NIM$_b$). We assume the NIM slabs to be isotropic and to have a 3D structure (as in \cite{bloemer2005}) although the experiment of Shelby \etal \cite{shelby2001} used a 2D structure. It was shown recently that a dielectric cavity formed by two such NIM slabs exhibits critical coupling near the edges of the stop band of NIM$_t$ \cite{sdgol2009}. The interplay of the FP resonances of the cavity with the stop band features of the NIM layers resulted in the critical coupling phenomenon. Further, varying the angle of incidence, which controls the width of the NIMs' stop gap, was used to achieve critical coupling at other frequencies. Recall that there is practically no reflection or transmission from a critically coupled cavity resulting in an almost perfect absorption of the incident energy by the structure. In this study our goal is to investigate the effects of cavity-atom interaction when the intracavity medium (i.e., the dielectric layer) is doped with resonant atoms. We assume the atomic medium to be nonmagnetic and its response to be given by the following dielectric function \\
\begin{equation}
\epsilon_2(f)= \epsilon_h+\frac{f_p^2}{f_{02}^2-f^2-i\gamma_2 f} \label{eq:eps2}~,
\end{equation}
where $\epsilon_h$ is the dielectric constant of the host material, $f_p^2$ is proportional to the dopant density, $f_{02}$ and $\gamma_2$ are the resonance frequency and the decay rate of the atom, respectively. Furthermore, the atomic resonance frequency is assumed to be close to the CC frequency of the undoped cavity. In the following section we demonstrate that such structures can lead to splitting of the CC dip for sufficient atom-cavity coupling determined by the dopant density. Such normal mode splittings have been studied in detail in the context of FP, modulated FP or spherical cavities (supporting the whispering gallery modes) \cite{MangaRao2004, sdgoptcomm93, sdgoptcomm92, sdgoptcomm95}. 
\par It is clear that the reflection and transmission profile for a system shown in figure \ref{fig:scheme} can be calculated using the standard characteristic matrix approach  \cite{yeh, bornwolf, sdg1998}. Both the reflection and transmission coefficients have a common denominator, the poles of which bear the information about the characteristic frequencies of the system. The corresponding dispersion relation can be written as \cite{sdgoptcomm93, sdg1998}
\begin{equation}
D=(m_{11}+m_{12}p_f)p_i + (m_{21}+m_{22}p_f) = 0~, \label{eq:disp}
\end{equation}
where $m_{ij}$ ($i,j=1,2$) are the elements of the total characteristic matrix of the structure,  $p_{i,f}=\sqrt{\epsilon_{i,f}/\mu_{i,f}} \cos{\theta_{i,f}}$ (for TE polarization) and $\theta_{i,f}$ are the angles of incidence and emergence in the first and the last medium, respectively. The dispersion relation  (\ref{eq:disp}) can be solved only for complex frequencies, which carries all the information about the split modes and their associated decay rates. In fact, the real part of the roots gives the locations of the split modes, while the imaginary part corresponds to the width of these resonances. It is understood that it is nontrivial to solve equation (\ref{eq:disp}) and obtain the different branches. We use a nonlinear root finder and present the results in the next section.
\section{Results}
In what follows we use scaled frequencies (in units of $f_0 = 10$GHz $=c/\lambda_0$) and lengths (in units of $\lambda_0$) to enable us to deal with dimensionless quantities (as in \cite{bloemer2005}) and also for reference. We use the same set of parameters as in the earlier work to achieve critical coupling with an undoped spacer layer \cite{sdgol2009}. The parameters for the lower NIM slab (NIM$_b$) are as follows: $f_{mp}=1.095$, $f_{ep}=1.28$, $f_{mo}=1.005$, $f_{eo}=1.03$, $\gamma=0.001$ and $d_3 = 5$, while for NIM$_t$ we take the same values, except for $f_{mp}=1.14$, $f_{ep}=1.175$  and $d_1=3$. This choice of parameters ensures that the stop band of NIM$_t$ lies inside that of NIM$_b$. The host material for the spacer layer has $\epsilon_h=3.8$. The entire layered structure is embedded in air ($\epsilon_i=\epsilon_f=1$, $\mu_i=\mu_f=1$). The spacer layer thickness ($d_2\approx 7$) is fixed to attain CC for normal incidence near the right edge of 
\begin{figure}\centering
\includegraphics[width=10cm]{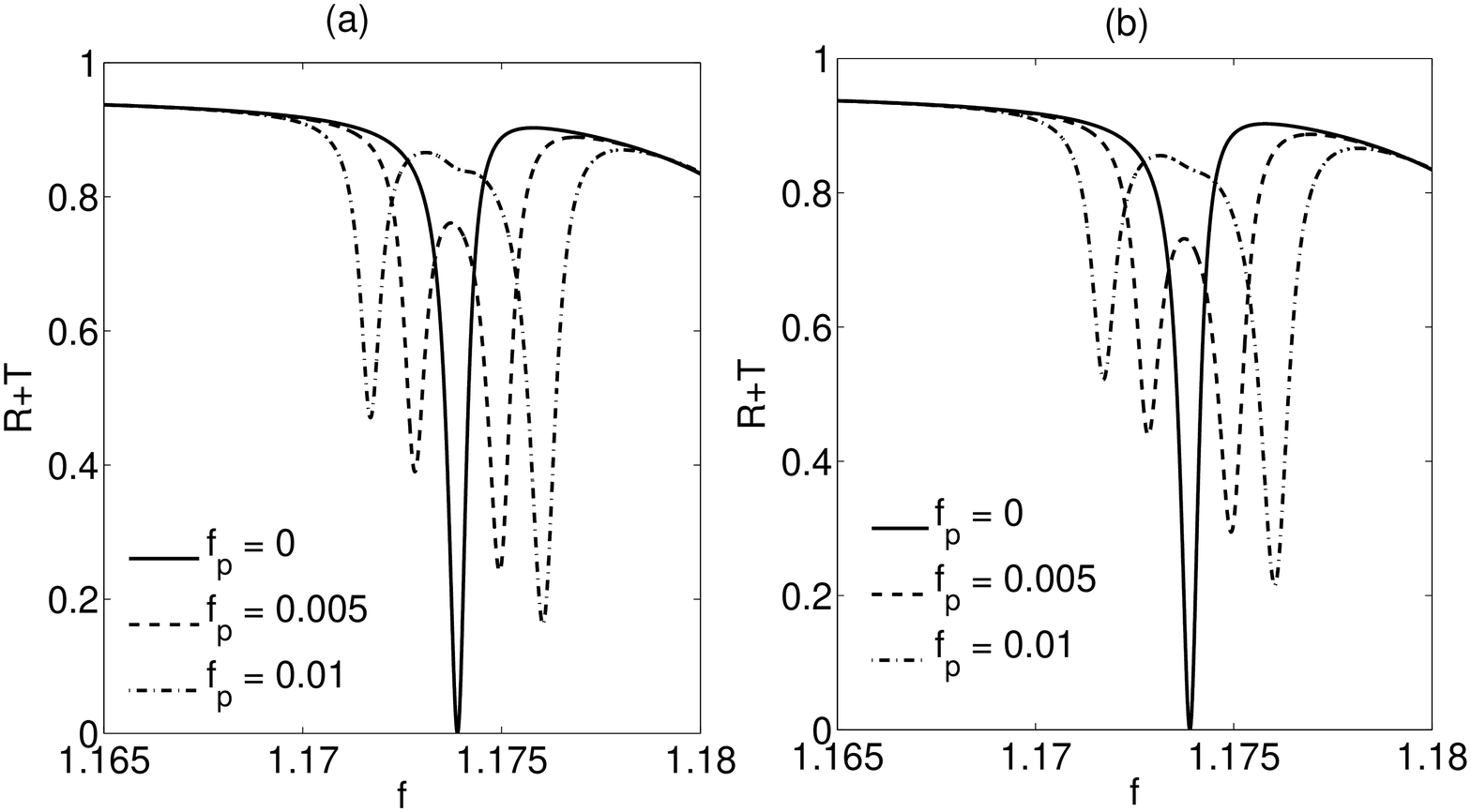}
\caption{Splitting of the CC resonance at normal incidence for two dopant atom decay rates (a) $\gamma_2=0.0008$  and (b) $\gamma_2=0.001$  with densities $f_p=$0.005 (dashed), 0.01 (dash-dot). The result for the undoped spacer layer ($f_p=0$) is reproduced (solid curve) for reference \cite{sdgol2009}. Here $d_1=3$, $d_2\approx 7$, $d_3=5$. $f_{02}\approx 1.174$ is near degenerate to the CC frequency (when $f_p=0$). Note that all frequencies are scaled in units of $f_0$.}\label{fig:2}
\end{figure}
\begin{figure}[b]\centering
\includegraphics[width=10cm]{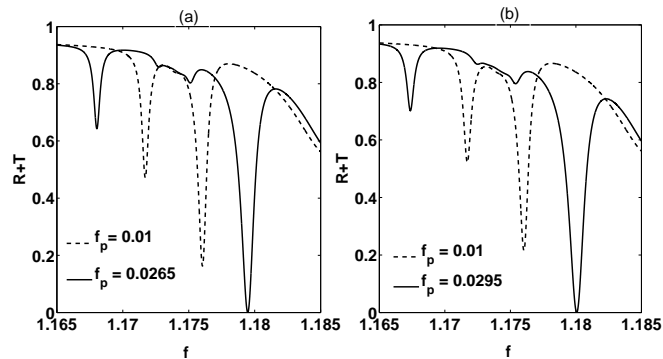}
\caption{CC is achieved (solid curves) for the split resonance at higher frequency ($f$) for two dopant decay rates (a) $\gamma_2=0.0008$, $f_p=0.0265$, $f=1.179$  and (b) $\gamma_2=0.001$, $f_p=0.0295$, $f=1.18$. The split resonances at $f_p=0.01$ (dashed curves) are produced for reference. The other parameters are the same as in figure \ref{fig:2}. }\label{fig:3}
\end{figure}
the stop band of NIM$_t$ (solid curves in figure \ref{fig:2}) without doping the cavity ($f_{p}=0$) \cite{sdgol2009}. The resonance frequency of the dopant atom is chosen to be near-degenerate (not identical) with this CC frequency at $f_{02}\approx 1.1738$ to ensure a strong interaction of the atom with the cavity resonance at CC. We study the total scattering ($R+T$) of the structure at normal incidence using the dopant density as a parameter. The calculations for total scattering have been performed for two dopant densities, namely, $f_p =0.005, 0.01$. The results for two decay rates ($\gamma_2 = 0.0008, 0.001$) are shown in figures \ref{fig:2}(a) and (b), respectively. These figures clearly demonstrate the splitting of the CC dip as a manifestation of the strong atom-cavity interaction (compare with the results of references \cite{sdgoptcomm93, sdgoptcomm95}). Unfortunately the critical coupling is lost at both the split frequencies. However, the higher frequency dip is closer to critical coupling for larger densities. Thus a larger dopant density can restore critical coupling at the right frequency component for a denser atomic system. This restoration is shown in figures \ref{fig:3}(a) and (b) for both the values of $\gamma_2$. For example, for $\gamma_2=0.0008$ ($\gamma_2=0.001$), CC is restored at $f_p=0.0265$ ($f_p=0.0295$). The dopant density may thus be used as a parameter to tune the CC frequency of the coupled system. A comparison of the left and right panels of figures 2 and 3 reveal that the splittings are more in case of lower atomic decay, which is quite expected. In fact the resolution of the split resonances depend on how the coupling compares with the decay rates. The lower the decay rates, the higher should be the splitting. Another pertinent feature that should be noted from these figures is that the width of the higher (lower) frequency split resonance increases (decreases) with increasing $f_p$. 
\begin{figure}[b]\centering
\includegraphics[width=10cm]{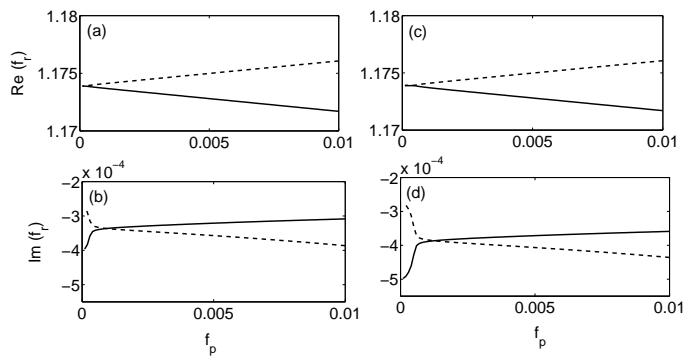}
\caption{The real ((a) and (c)) and imaginary ((b) and (d)) parts of the roots of the dispersion relation ($f_r$) of the layered structure as a function of the atomic density at $\gamma=0.0008$ (left panel) and $0.001$ (right panel). The solid (dashed) curve corresponds to the split resonance at the lower (higher) frequency. }\label{fig:4}
\end{figure}
\par 
We now show that all the above features can easily be understood based on an analysis of the roots of the dispersion relation (\ref{eq:disp}). The real and imaginary parts of the roots (as functions of the dopant density) for the two values of $\gamma_2$ ($\gamma_2=0.0008$, $0.001$) are shown in the left and right panels of figure \ref{fig:4} . The solid and dashed lines give the two branches for both. The well known frequency and lifetime splittings \cite{sdgoptcomm93, sdgoptcomm95} can easily be read from the top and bottom panels of figure \ref{fig:4}. It is clear that the split mode frequencies move away from each other with increasing $f_p$ (top panel of figure \ref{fig:4}) in conformity with the previous total scattering results. The split resonance at the higher frequency (dashed curves) widens, whereas, the one at lower frequency (solid curves) becomes narrower with increasing $f_p$ (see the bottom panels in figure \ref{fig:4}). 
\begin{figure}\centering
\includegraphics[width=10cm]{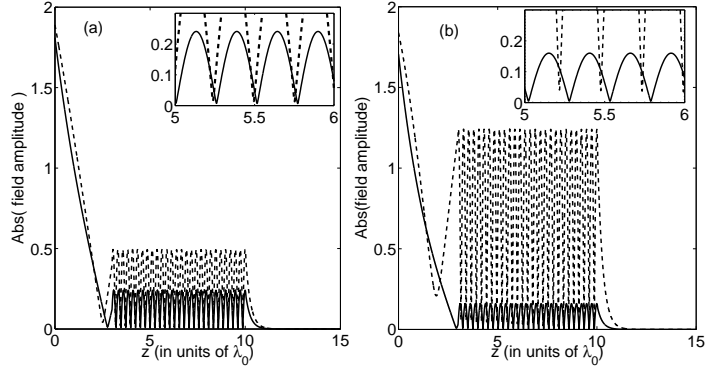}
\caption{The field distribution inside the structure for the parameters in figure \ref{fig:3}(a). (a) $f_p=0.01$ with split resonances at $f=1.172$ and $1.176$ (b) $f_p=0.0265$ with split resonances at $f=1.168$ and $1.179$ . The solid(dashed) curves correspond to the lower(higher) frequency split resonance. $z=0$ corresponds to the air-NIM$_t$ interface and $z=d_1 + d_2 +d_3 \approx 15$ to the NIM$_b$-air interface. The insets are enlarged portions of the corresponding plots.}\label{fig:5}
\end{figure}
\par 
The correspondence of the split resonances to symmetric and anti-symmetric modes, as in vacuum Rabi splitting, can be inferred from a study of the field distribution inside the structure. The absolute value of the field amplitude must reach zero inside the cavity for the antisymmetric mode but not for the symmetric one. We use the parameters in figure \ref{fig:3}(a) and plot in figure \ref{fig:5} the absolute value of the amplitude at both the left (antisymmetric) and right (symmetric) resonances (solid and dashed lines, respectively). $z$ in figure \ref{fig:5} corresponds to the depth inside the layered structure from the air-NIM$_t$ interface. From the insets in figure \ref{fig:5} it is seen that the magnitude of the field amplitude touches zero for the antisymmetric mode only. Further, we note the significant field enhancement in case of the critically coupled cavity (see figure \ref{fig:5}b)
\section{Conclusion}
In conclusion, we have studied a critically coupled cavity and shown that the critical coupling dip can exhibit a normal mode splitting when the intracavity medium is doped with resonant atoms. The intracavity medium with dopant atoms was modelled by a Lorentz type dielectric with characteristic resonance frequency and decay rate. Numerical results for the total scattering R+T was obtained using a characteristic matrix formulation. The corresponding dispersion relations were solved for the location and width of the split resonances. Moreover, it was demonstrated that the atomic dopant density can be used as an additional handle to achieve critical coupling with respect to one of the split resonances. 

\end{document}